\documentclass[twocolumn]{article}  
\usepackage{graphics}
\usepackage{epsf}

\def\bra#1{\left\langle\,#1\,\right|} 
\def\ket#1{\left|\,#1\,\right\rangle}
\begin{document}

\title{Stable nondegenerate optical parametric oscillation at degenerate frequencies in Na:KTP}
\author{Sheng Feng and Olivier Pfister\thanks{Corresponding author. opfister@virginia.edu}\\\ \\
\em Department of Physics, University of Virginia, \\ \em 382 McCormick Road, Charlottesville, VA 22904-4714, USA}
\maketitle 
\begin{abstract}
We report the realization of a light source specifically designed for the generation of bright continuous-variable entangled beams and for Heisenberg-limited inteferometry. The source is a nondegenerate, single-mode, continuous-wave optical parametric oscillator in Na:KTP, operated at frequency degeneracy and just above threshold, which is also of interest for the study of critical fluctuations at the transition point. The residual frequency-difference jitter is $\pm$ 150 kHz for a 3 MHz cold cavity half-width at half maximum. We observe 4 dB of photon-number-difference squeezing at 200 kHz. The Na:KTP crystal is noncritically phase-matched for a 532 nm pump and polarization crosstalk is therefore practically nonexistent.
\end{abstract}

\section{Introduction}

Optical parametric oscillators (OPO's) are well known sources of nonclassical light. Type-I and type-II OPO's have been used to generate single-mode \cite{smsq} and two-mode \cite{ou,tb} squeezed states and to implement the Einstein-Podolsky-Rosen (EPR) paradox \cite{epr} experimentally \cite{ou}, as proposed in \cite{reid,reid2}. The latter has become increasingly important as continuous variables (CV) have been proposed and used to implement quantum information protocols \cite{vaidman,braunstein} such as quantum teleportation \cite{furusawa} and quantum dense coding \cite{peng}. This paper is focused on CV and OPO's, but one should not forget, of course, that single-photon coincidence detection coupled with spontaneous parametric downconversion \cite{kwiat} has played a major role in the violation of Bell's inequalities \cite{bi} (practically unfeasible with CV), and the realization, for example, of teleportation \cite{z1}, multipartite entanglement \cite{z2}, and very recently quantum computer gates \cite{franson}.

Going back to CV, the generation of CV-EPR states requires simultaneous squeezing of two compatible joint operators made from noncompatible local operators \cite{epr,vaidman}, or physical variables, such as number sum (difference) and phase difference (sum) ($N_\pm,\phi_\mp$), or amplitude-quadrature sum (difference) and phase-quadrature difference (sum) ($X_\pm,P_\mp$). Reid and Drummond proposed to use above- and below-threshold OPO's in order to squeeze $N_-$ and $\phi_+$ \cite{reid,reid2}. In the first experimental implementation of the EPR paradox \cite{ou}, a nondegenerate OPO below threshold was used. An interesting variant also demonstrated was to use two degenerate OPO's below threshold, each squeezing a complementary quadrature, and to overlap their fields using a balanced beam splitter \cite{furusawa}. As pointed out by Lugiato et al., the unitary transformation describing the beam splitter transforms the nondegenerate and degenerate Hamiltonians into one another \cite{lugiato}. Several related entanglement experiments have been recently reported \cite{peng2,lam,polzik}, and one ought also to mention the use of third-order nonlinearity in fibers to generate bright EPR beams without the use of OPO's \cite{leuchs}.

It is interesting to note that, despite the fact that the initial proposal for experimentally realizing the EPR paradox was to use an OPO above threshold \cite{reid}, this particular implementation has not yet been carried out. Rather, in all EPR implementations to date \cite{ou,furusawa,peng2,lam,polzik}, OPO's have been used exclusively below threshold, as parametric amplifiers of either vacuum or bright modes. The reason for this choice is that the proposal of Ref.~\cite{reid} is very challenging. It requires either a triply resonant OPO with higher finesse for the pump than for the signal and idler, which presents technical difficulties, or a doubly resonant resonant OPO just above threshold. Moreover, in both cases, the OPO must be emitting close-to-degenerate signal and idler frequencies, which is difficult to achieve because the well-known clustering effect in a doubly resonant type-II OPO puts stringent stabilization requirements on the crystal temperature, the cavity length, and the pump frequency. One has thus to address two experimental challenges, using a doubly resonant OPO: to achieve stable oscillation on the frequency degenerate mode and do so close to the threshold, where the output power is the least stable. 

We have solved these two issues and present, in this paper, the experimental realization of a stable type-II OPO, emitting at frequency degeneracy while pumped only a few percent above threshold. Note that stable twin beams have been achieved before, in particular in the record-breaking number-difference squeezing experiments by the group of Fabre and Giacobino \cite{tb} and in the more recent work of several other groups \cite{opo}, but our present work is, to the best of our knowledge, the first time that stable twin beams are obtained at precise frequency degeneracy and just above threshold.

Besides bright EPR beam generation, such a source is also extremely interesting for the study of critical fluctuations at the transition point, including possible generation of macroscopic entanglement \cite{drummond}, and, last but not least, for Heisenberg-limited interferometry (HLI). Recall that, if $N$ particles are sent into an interferometer and subsequently detected or measured, the usual phase sensitivity limit is the input beam splitter's shot noise limit $N^{-1/2}$ \cite{caves}, whereas the ultimate limit is the Heisenberg limit $N^{-1}$ \cite{caves2,ou2}. This is a general statement for any boson field. Examples of beam splitters are a half-reflecting mirror in photon optics and a $\pi/2$ resonant laser pulse in atom optics. Reaching the Heisenberg limit requires suppressing vacuum fluctuations at the input beam splitter of the interferometer \cite{caves2}, which was experimentally demonstrated using vacuum squeezing \cite{hli}. Holland and Burnett \cite{holland} also showed that one can reach the Heisenberg limit by using a source of indistinguishable twin modes, i.e.\ an input density matrix of the form $\rho=\sum_{n,m}\rho_{nm}\ket{n\,n}\bra{m\,m}$.   This was demonstrated using two single-mode amplitude-squeezed beams in Ref.~\cite{leuchs}, and also a pair of trapped ions  \cite{wineland}. Progress has also been made towards realizing HLI with Bose-Einstein-condensate Fock states \cite{kasevich}. It is interesting to note, however, that the common-mode statistics $(\rho_{nm})_{n,m}$ do not play any role in the phase sensitivity \cite{kim}, hence generating twin Fock states e.g.\ $\ket{n\,n}$ is {\em not} necessary. Therefore, an OPO emitting intense frequency-degenerate twin beams is an ideal candidate system for HLI, as it would bring, besides the required $N_-$ squeezing, the added benefit of a a narrow linewidth CW source, suitable to perform ultra-precise measurements such as gravitational-wave detection. Note also that, since HLI requires large photon numbers, it presents one fewer challenge than bright EPR state generation.

In the next section, we describe our experimental setup. We subsequently detail our experimental characterization of the source.

\section{Experimental setup}

\subsection{OPO crystal and cavity}

The Na:KTP crystal noncritically phase-matches type-II degenerate parametric downconversion from 532 to 1064 nm wavelengths along the $X$ axis at room temperature. Such a configuration practically annihilates polarization crosstalk while allowing the convenient use of highly stable Nd:YAG lasers. We use a $\rm 3\times 3\times 10\ mm^3$ prototype Na:KTP crystal, fabricated by Crystal Associates. The crystal is cut along its $X$ axis, and AR-coated at 1064 nm and 532 nm. It is placed inside an oven made of Oxygen-free, high-conductivity Copper, whose temperature is controlled to less than a millidegree by way of a servo loop. We used two kinds of loop filters with equal success, one home-made and one commercial from Wavelength Electronics. Tight temperature control of the nonlinear crystal is paramount in a type-II OPO because of the different temperature dependences of the indices of refraction for the two cross-polarized resonant signal and idler waves. 

The OPO has a standing-wave cavity formed by two plano-concave mirrors that have a radius of curvature of 5 cm and are separated by approximately 10.2 cm, thereby realizing a concentric resonator. One of the mirrors is mounted on a piezoelectric transducer (PZT). The cavity waist size is in between the Boyd-Ashkin confocal-focusing value $w_o(BA)=\sqrt{\ell\lambda/2\pi n}=31\ \mu$m and the Boyd-Kleinman optimum $w_o(BK)=0.59\, w_o(BA)=18\ \mu$m \cite{byer}, which is confirmed by OPO threshold measurements. The interferometer structure is built in super-Invar, which confers outstanding mechanical stability along with no visible thermal expansion effects. This makes for an ultrastable optical resonator that never needs to be realigned, the OPO threshold staying at its initial level for months on.

The concentric geometry is more delicate to work with than a plano-concave cavity (as used for example in hemilitic OPO's) since its spherical symmetry allows for many different cavity axes and does not constrain the alignment very strongly. This is, however, a conscious design choice as it allows us to easily find a cavity mode that coincides with the crystal's $X$ axis so as to completely eliminate walkoff. The OPO alignment procedure consists of aligning the resonator without the crystal first, and then inserting the crystal and orienting it between crossed polarizers in order to align the $X$ axis with the cavity axis. The cavity is then used as a resonant frequency doubler in order to optimize the crystal temperature for maximum SHG efficiency, which corresponds to the lowest OPO threshold. The SHG pump laser at 1064 nm (Lightwave Electronics 126-700-1064) is, once frequency-doubled, heterodyned with the OPO pump laser (Lightwave Electronics 142) and the resulting beat note at their frequency difference is tuned to zero. The reflectivities of the OPO mirrors are $R(1064$ nm$)=0.999$ and $R(532$ nm$)=0.6$ for the input coupler, and  $R(1064$ nm$)=0.990$ and $R(532$ nm$)=0.1$ for the output coupler. The OPO is therefore essentially doubly resonant for the signal and the idler.

\subsection{OPO stabilization}

\begin{figure}[ht]
\begin{center}
\resizebox{3in}{!}{\epsfbox{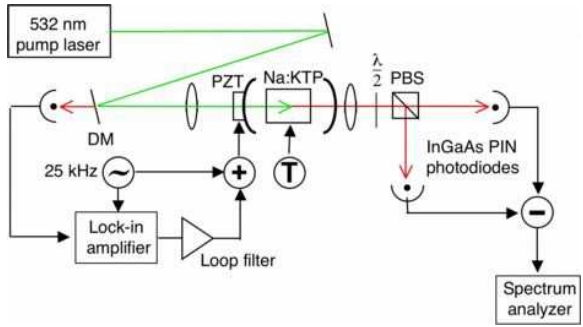}}
\end{center}
\caption{Experimental setup. DM: Dichroic mirror (reflects 532 nm; transmits 1064 nm).}
\label{setup}
\end{figure}

In order to ensure single-mode oscillation within the dense cluster structure (see below), we use standard laser stabilization techniques \cite{pdh}, which permit cavity-length control at the femtometer level \cite{hall}. The OPO cavity is electronically stabilized by two servo loops, one for temperature and one for cavity length (Fig.~\ref{setup}). 
The cavity-length error signal is picked off the small leak through the HR input coupler. A sinusoidal voltage at 25 kHz excites the first mechanical resonance of the PZT-mirror system and serves to modulate the OPO frequency. The weak leak signal is demodulated by a lock-in amplifier, processed by a loop filter, and finally applied to the OPO PZT. The good signal-to-noise ratio of this method allows us to operate with the OPO a few percent above threshold, which is of interest for quantum optics with bright beams as it will allow us to easily perform homodyne squeezing measurements with a stronger local oscillator, something believed difficult \cite{peng2} because of the current power-withstanding limitations of high-efficiency photodiodes.

\subsection{Detection}

The OPO signal and idler beams are collimated by an AR-coated lens and then traverse a variable beam splitter comprised of a half-wave plate followed by a low loss polarizing beam splitter ($T_p,R_s>0.99$). 
\begin{figure}[ht]
\begin{center}
\resizebox{3.25in}{!}{\epsfbox{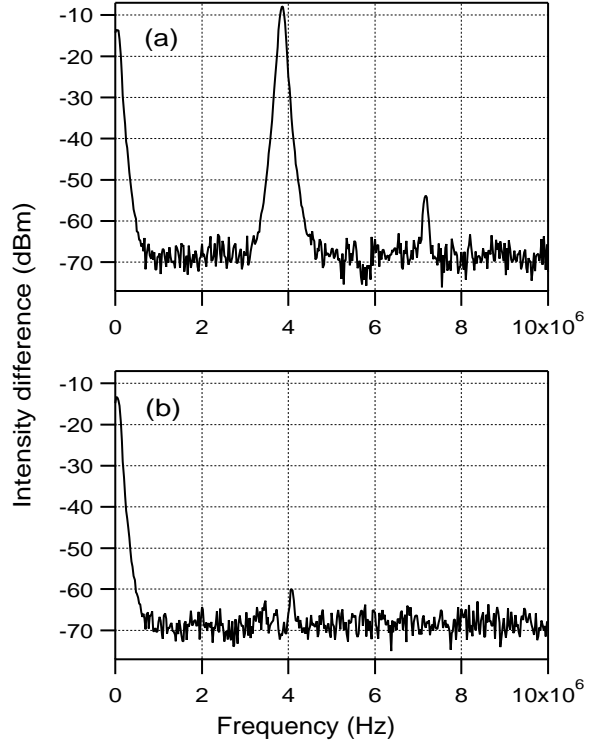}}
\end{center}
\caption{\label{polcrosstalk} (a): Intensity difference spectrum of twin beams through a balanced beam splitter (half-wave plate @ $\pi/8$ rad). The 4 MHz signal-idler beat note is clearly visible, along with its second harmonic (preamplifier distortion). (b): Intensity difference spectrum of twin beams (half-wave plate @ $0$ rad). The residual 4 MHz beat note is 52 dB below its maximum level of fig.~\ref{polcrosstalk}(a). (No intensity squeezing is visible here because the OPO beams are attenuated before detection.) Resolution bandwidth: 100 kHz. Single sweeps.}
\end{figure}
We determine the reference angle of the wave plate axes $\alpha=0$ by minimizing the beat note on the detectors, which clearly separates the twin beams. For other angles relative to this one, the beams are thus mixed in the equivalent way to a beam splitter with reflectivity $R=\cos^2(2\alpha)$. When $\alpha=\frac\pi 8$, the setup is equivalent to a balanced beam splitter. This arrangement also allows us to check the presence of polarization crosstalk, i.e.\ the existence of a residual beat note for  $\alpha=0$. As seen in Fig.~\ref{polcrosstalk}, this effect is negligible. 

The detection is performed by two high efficiency ($>95\%$) InGaAs photodiodes (JDS Uniphase ETX500T) whose photocurrents are subtracted and sent into an ultralow-noise transimpedance preamplifier. We measure the common-mode rejection ratio to be 42 dB @ 100 kHz, which enables us to have a shot-noise limited detection bandwidth as low as 200 kHz.

\section{Experimental results}

\subsection{Intensity-difference squeezing}

\begin{figure}[ht]
\begin{center}
\resizebox{3.25in}{!}{\epsfbox{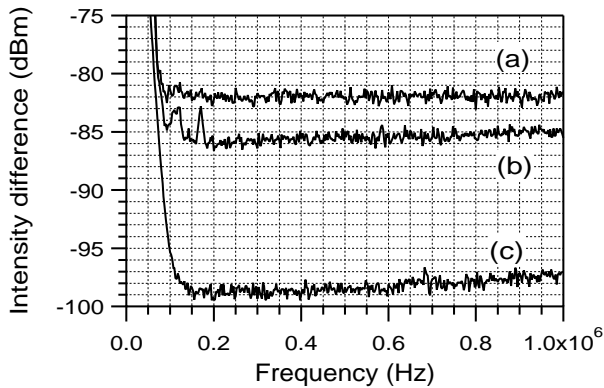}}
\end{center}
\caption{OPO  intensity-difference squeezing spectrum. (a): shot noise level --- two-detector difference signal in beam splitter mode (half-wave plate @ $\pi/8$ rad). (b): number-difference squeezing --- two-detector OPO difference signal (half-wave plate @ $0$ rad). (c): electronic detection noise (no light). Resolution bandwidth: 10 kHz. 100-sweep averages.}
\label{sq}
\end{figure}

Figure \ref{sq} displays the measured photocurrent spectrum for different cases, with a maximum measured squeezing of 4 dB @ 200 kHz. 
This detection bandwidth is quite low and unusual for squeezing experiments (but see also Ref.~\cite{bowen}). Reaching even lower quantum-limited detection bandwidths (1 kHz or less) would be of tremendous interest for HLI applied to gravitational-wave detection. The squeezing spectrum has a maximum of 4 dB and an inverted Lorentzian shape of width the cold cavity linewidth \cite{tb}, which is 3 MHz here. Several straightforward measures can be taken to improve on this preliminary result, such as removing the photodetectors' uncoated windows, using a shorter crystal, and using ultralow-loss OPO mirrors \cite{rempe}. The focus of this paper, however, is not as much on squeezing as on stability.

\subsection{Frequency control}

It is well-known that, in a doubly resonant type-II OPO, the double resonance condition for the cross-polarized signal and idler inside a common cavity containing a birefringent nonlinear crystal leads to a densely clustered mode structure \cite{shen}. This leads to a several hundred-fold increase of the constraints for length stability of the cavity, and demands that an electronic servo-lock loop be used. Theoretical analyses of the type-II OPO have been carried out in Refs.~\cite{deb,wong} so we just outline the relevant points here. The double resonance condition for the signal and the idler is $L+n_{s,i}\ell=p_{s,i}\lambda_{s,i}/2$, $L$ being the cavity length outside the crystal, $\ell$ the crystal length, $n_{s,i}$ the refraction indices of the signal and idler waves, $\lambda_{s,i}$ their respective wavelengths, and $p_{s,i}$ their respective mode numbers (positive integers). The equivalent resonance conditions for the optical frequencies $\nu_{s,i}=c/\lambda_{s,i}$ are $\nu_{s,i}=p_{s,i}\Delta_{s,i}=p_{s,i}c/[2(L+n_{s,i}\ell)]$, where $\Delta_{s,i}$ are the free spectral ranges (FSR) of the signal and the idler. The resonance frequency change for a small air path change, which corresponds to actuating one of the mirrors with a PZT, is therefore
\begin{equation}\label{tune}
\frac{d\nu_{s,i}}{dL} = \Delta_{s,i}\left(\frac{dp_{s,i}}{dL} - \frac 2{\lambda_{s,i}}\right).
\end{equation}
From this equation, one finds the well-known minimum length variation $\delta L=\pm\lambda/2$ for which a {\em mode hop} ($\delta p=\mp 1$, $\delta\nu=0$) occurs, in the case of a singly-resonant OPO or laser. In a doubly-resonant OPO, the additional constraint that both frequencies sum up to the pump frequency $\nu_p=\nu_s+\nu_i$ modifies the length tuning of the OPO cavity in a dramatic way, as the OPO will not lase for all combinations of the signal and idler modes. The result is a tightly packed cluster structure. For a stable and narrow pump frequency, which is the case here, one has $d(\nu_s+\nu_i)/dL = 0$. The length displacements corresponding to mode hops are then given by, to first order in the birefringence $\delta n=n_s-n_i$,  
\begin{equation}
\delta L \simeq \left[ (\delta p_s+\delta p_i) + \frac 1 2\, \frac{\delta n}{\bar n + L/\ell}\ (\delta p_s-\delta p_i)\right]\,\frac{\lambda_p}2, 
\end{equation}
\begin{figure}[ht]
\begin{center}
\resizebox{3.25in}{!}{\epsfbox{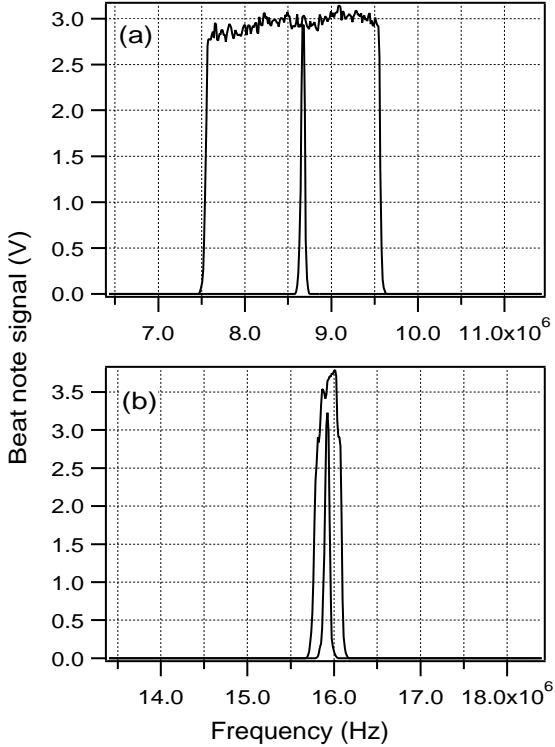}}
\end{center}
\caption{\label{jitter} OPO beat note spectrum, 5 MHz span.  The narrow peaks are single spectrum analyzer sweeps (14 ms), and the broad features are the corresponding ``max hold" traces for one minute or more. (a): Free-running OPO 4 times above threshold. The HWHM of the narrow peak is 50 kHz and the drift range is 2 MHz. Free-running oscillation stays single-mode without servo because the OPO mode HWHM is of the order of 10 MHz (the cluster modes overlap. See text). (b): Frequency-locked OPO four times above threshold. The narrow HWHM is 60 kHz and the drift range is 310 kHz, only limited by the signal-to-noise ration of the servo error signal. Note that the fast beat-note linewidth is limited by the 30 kHz resolution bandwidth of the analyzer.}
\end{figure}
where $\lambda_p$ is the pump wavelength, $\bar n=(n_s+n_i)/2$, and $\delta p_{s,i}$ are integers. Mode clusters are labeled by $(\delta p_s+\delta p_i)$ and separated by half a pump wavelength. Inside a given cluster, the modes are labeled by $(\delta p_s-\delta p_i)$ and separated by ($\delta p_s=-\delta p_i=\pm1$)
\begin{equation}
\delta L_{min} \simeq \frac{\delta n}{\bar n+ L/\ell}\ \frac{\lambda_p}2 \ll \lambda_p,
\end{equation}
the signal-idler frequency difference or beat note $\nu_b=\nu_s-\nu_i$ differing, between two consecutive modes, by the sum of the FSR's. In the case of our experiment, $\bar n= 1.8$, $\delta n=0.09$, $L=9.2$ cm, and $\lambda_p=532$ nm, give clusters every 266 nm with mode hops
\begin{equation}
\delta L_{min}\simeq\frac{\lambda_p}{244}\simeq 2.2\rm nm. 
\end{equation}
We now derive the tuning coefficients of the OPO sum and difference frequencies $\nu_\pm=\nu_s\pm\nu_i$ with respect to the pump frequency $\nu_p$, cavity length $L$, crystal temperature $T$, and voltage across the crystal  $V$ (electro-optic tuning). Using Eq.~(\ref{tune}) and analogs and assuming $\nu_-=0$ and no mode hops, one finds
\begin{eqnarray}
\left(\frac{\partial\nu_\pm}{\partial L}\right)_{T,V,\nu_p}  &=& (-5.11,-0.02)\ \mathrm{MHz/nm}\label{Lcoeff};\\
\left(\frac{\partial\nu_\pm}{\partial T}\right)_{L,V,\nu_p}  &=& (-2.12, 0.24)\ \mathrm{MHz/mK};\\
\left(\frac{\partial\nu_\pm}{\partial V}\right)_{L,T,\nu_p}  &=& (1.34, 0.59)\ \mathrm{MHz/V}\label{Vcoeff};\\
\left(\frac{\partial\nu_\pm}{\partial\nu_p}\right)_{L,T,V} &= & (1,0).
\end{eqnarray}
The next question is: what are the sources of noise? In the experiment, the pump frequency noise is $<10$ kHz/ms with a slow thermal drift $<10$ MHz/min; the temperature servo gives $\delta T<1$ mK in a 1 Hz bandwidth; and the cavity servo gives $\delta L = 0.01$ nm in a 3 kHz bandwidth. No electro-optic servo is applied as the noise on $V$ is assumed negligible. The results of these stabilization efforts are displayed on fig.~\ref{jitter},
which shows recordings of the beat note at two very different time scales, with the temperature controlled and a free-running and locked cavity. The measured fast linewidths are of the order of 50 kHz for a 30 kHz resolution bandwidth, and are thus instrument-limited. The locked residual jitter linewidth is 310 kHz, down from the free-running range of 2 MHz, and the jitter frequency is 100-500 Hz. This 310 kHz residual linewidth on the beat note is larger than expected from the length and temperature tuning coefficients calculated above, and is too fast for being imputed to residual temperature fluctuations. We deduce that the beat note stability is degraded by other factors. 

In fact, we have observed that the crystal is slightly inhomogeneous, likely due to a dopant concentration gradient. As a result, moving the crystal transversally in the beam tunes its birefringence and therefore the OPO, which, therefore and unexpectedly, becomes extremely sensitive to table vibrations {\em transverse} to its axis. We have confirmed this by observing that the jitter's spectral signature matches that of the table vibration measured with accelerometers, both sharing in particular sharp resonances at 70 and 100 Hz. This produces noise on $\nu_\pm$, and Eq.~(\ref{Lcoeff}) shows that the cavity length servo is ill-equipped to do much correction on $\nu_-$ because of the low tuning coefficient. As already mentioned, the temperature servo is too slow to be useful here, hence further narrowing of the beat note spectrum will require an electro-optic servo on the crystal so as to make use of the favorable tuning coefficient of Eq.~(\ref{Vcoeff}).

Note, however, that the current level of performance is already outstanding and satisfactory for EPR and HLI experiments, the beat note being tunable to exactly zero hertz (within the residual jitter range) by a simple temperature adjustment.

\subsection{Intensity control}

\begin{figure}[ht]
\begin{center}
\resizebox{3.25in}{!}{\epsfbox{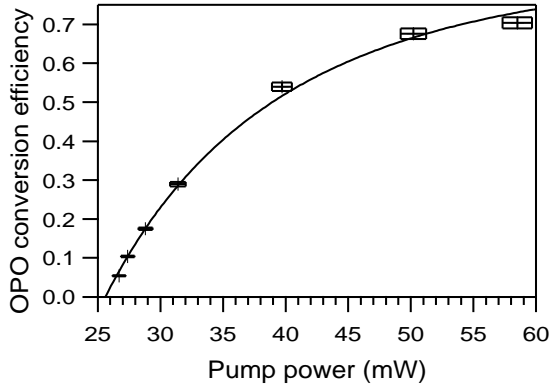}}
\end{center}
\caption{Measurement and fit of the conversion efficiency of the OPO locked on the degenerate mode. $\rho=\frac K N (\sqrt N -1)$, where $N=P_{p}/P_\mathit{th}$. Fit parameters are $P_\mathit{th}=25.6(2)$ mW, $K=3.26(6)$, $\chi^2=1.5\ 10^{-3}$.}
\label{eff}
\end{figure}

In order to determine how close to the threshold the OPO can stably be operated, which is crucial in order to determine whether EPR correlations can be observed, we have investigated the conversion efficiency given by \cite{byer}
\begin{equation}\label{conveff}
\rho = \frac{P_{out}}{P_p} = \frac{P_s+P_i}{P_p} = \frac K N (\sqrt N -1),
\end{equation}
where $N=P_p/P_{th}$ is the number of times above threshold (and should be as close to 1 as possible) and $K$ is a constant whose value is 2 for a standing wave cavity with a single-pass pump and up to 4 if the pump is reflected, the value of 4 corresponding to a ring resonator or to a double-pass pump with the right phase relationship with the signal and idler fields $\delta\phi=\phi_p-\phi_s-\phi_i=-\pi/2$ \cite{shen}. The value of $K$ therefore depends on the differential phase shift imparted to $\delta\phi$ by reflection on the OPO output mirror and on air dispersion. We measured the OPO output power for different values of the pump power for the frequency-degenerate mode in stably locked conditions. Figure \ref{eff} displays a plot of $\rho$ versus $P_p$, as well as a least-squares fit using Eq.~(\ref{conveff}). Since we do not know the mirror's phase shift, we leave $K$ as a free parameter in the fit. 
The lowest point on the plot corresponds to $N=1.04$, $\rho=5\%$, and $P_{out}= 700\ \mu$W per beam. We have in fact observed stable oscillation for $N$ as low as $1.01$, i.e.\ output powers of a few hundred microwatts per beam. This should indeed be in the regime where EPR correlations are expected \cite{reid}. The OPO output noise on each individual beam is mainly that of the pump laser, i.e.\ fairly low (but with relaxation oscillations) and is shot-noise limited above 1.5 MHz. As already seen on fig.~\ref{sq}, intensity-difference measurements are shot-noise limited above 200 kHz.  This speaks to the low level of classical intensity noise. Figure \ref{jitter} (a) also gives an idea of longer-term intensity fluctuations of the {\em unlocked} OPO in a single-detector measurement.

The theoretical threshold is 12 mW \cite{byer}. We measured about 15 mW for the strongest cluster mode and 25.6(2) mW for the degenerate mode, which is not the strongest (center) cluster mode. This is due to a slight misalignment of the phase-matching angle of the crystal. Correcting it is quite doable but is always a delicate operation as it couples to the cavity alignment. We observe that the power level of the degenerate mode remains remarkably constant over time, even with the experiment (including the pump laser) turned off each night, testifying to the stability of the optical alignments. 

It is possible to achieve even lower powers with the same threshold, but the capture range of the servo decreases as the gain linewidth and this degrades the locking performance. We are currently implementing lower-noise detection techniques aiming at increasing the signal-to-noise ratio by one order of magnitude and at expanding the capture range of the servo. One should also note that, since the error signal is mixed with a considerable amount of vacuum fluctuations (99.99\%), we only use a classical signal to lock. As a consequence, our servo cannot correct quantum fluctuations, be they phase difference noise or critical threshold fluctuations, which will allow us to observe them without interference from the locking electronics.

\section{Conclusion}

We have experimentally demonstrated the stable operation of a frequency degenerate type-II OPO above threshold, with negligible polarization crosstalk, using a noncritically phase-matched Sodium-doped KTP nonlinear crystal. The degenerate mode is obtained reproducibly without the need to realign the OPO cavity after the initial alignment, and its intensity is controlled as low as just a few percent above threshold. The twin beams exhibit 4 dB of number-difference squeezing at 200 kHz, which illustrates the low classical noise of our setup. Pushing this limit to even lower frequencies, below 1 kHz, is important from the perspective of applying HLI to the detection of gravitational waves. All stability and squeezing performances can be significantly improved on this starting point, which is already suitable for the generation of bright EPR beams and HLI.

\section*{Acknowledgments} 

We thank Ken Nelson for his invaluable help and advice with high-performance electronics design, Harvey Sugerman for building some of the electronics, Richard Stolzenberger for help with Na:KTP, and Rodger Ashley and Roger Morris for machining our super-Invar resonator structure.

This work is supported by the U.S. Army Research Office and has been supported by the Jeffress Trust Foundation. 


\end{document}